\documentclass[
    ,final            
  ,numberedheadings 
  ]
  {aipproc}

\layoutstyle{6x9}
\def\msun{$M_{\odot}$}
\def\PR{{\em Phys. Rev. D }}

\begin{document}

\title{Extracting Information about EMRIs using Time-Frequency Methods}


\author{Linqing Wen }{
  address={Max Planck Institut fuer Gravitationsphysik, Albert-Einstein-Institut Am Muehlenberg 1,  D-14476 Golm, Germany}}

\author{Yanbei Chen}{
  address={Max Planck Institut fuer Gravitationsphysik, Albert-Einstein-Institut Am Muehlenberg 1,  D-14476 Golm, Germany}
}

\author{Jonathan Gair}{
  address={Institute of Astronomy, University of Cambridge, Madingley Road, Cambridge, CB3 0HA, UK}
 }

\begin{abstract}
The inspirals of stellar-mass compact objects into supermassive black holes are some of the most exciting sources of gravitational waves for LISA.  Detection of these sources using fully coherent matched filtering is computationally intractable, so alternative approaches are required. In \citep{wen05a}, we proposed a detection method based on searching for significant deviation of power density from noise in a time-frequency spectrogram of the LISA data. The performance of the algorithm was assessed in \citep{jon05} using Monte-Carlo simulations on several trial waveforms and approximations to the noise statistics. We found that typical extreme mass ratio inspirals (EMRIs) could be detected at distances of up to 1--3 Gpc, depending on the source parameters.  In this paper, we first give an overview of our previous work in \cite{wen05a, jon05}, and discuss the performance of the method in a broad sense.  We then introduce a decomposition method for LISA data that decodes LISA's directional sensitivity. This decomposition method could be used to improve the detection efficiency, to extract the source waveform, and to help solve the source confusion problem.  Our approach to constraining EMRI parameters using the output from the time-frequency method will be outlined.
 
\end{abstract}

\maketitle


\section{Background}
Astronomical observations indicate that many galaxies host a supermassive black hole (SMBH) in their center. The inspirals of stellar-mass compact objects into such SMBHs with mass $M\sim\mbox{few}\times10^{5}M_{\odot}$--$10^{7}M_{\odot}$ constitute one of the most important gravitational wave (GW) sources for the planned space-based GW observatory LISA. Typical EMRI events will come from objects on elliptical orbits around spinning black holes and this leaves a significant imprint on the emitted GWs. The parameter space of possible GW waveforms is large since the masses of the two components, the amplitude and relative direction of the SMBH spin, the orbital pericenter and eccentricity and the object's initial phase relative to pericenter all need to be specified.   

EMRI waveforms are complicated.  There are three characteristic frequency components arising from the orbital motion, the nodal precession due to spin-orbit coupling and general relativistic perihelion precession. The GW power is therefore spread over a wide frequency range at harmonics of these three major frequencies. Over an observation lasting LISA's nominal lifetime of three years, there is also significant time evolution of these frequency components and variation in their relative amplitudes~\cite{leor04}. Moreover, due to LISA's motion around the sun, the detector's response to GWs undergoes a time-varying amplitude modulation and frequency Doppler shift that depends on the source direction.  The signal-to-noise ratio (SNR) of a ``typical'' EMRI event (e.g., a $10+10^6$ \msun\  system at 1 Gpc) is large if we can perform an optimal search using coherent matched filtering. However, the signal power in each Fourier frequency bin is very small, given the signal's large frequency spread over $N \sim Tf \sim 10^5$ bins.    

Study has shown that detection of EMRIs using the optimal fully coherent matched filtering is computationally impossible \citep{jon04}, so alternative search techniques are required.  A semi-coherent matched filtering algorithm has been proposal as an alternative \citep{jon04}. This involves matched filtering on short segments of the data, that are as long as is allowed using reasonable computational resources. The power in these segments is then added incoherently over the entire data stretch. Preliminary results for this search \citep{jon04}  suggest that the LISA EMRI detection rate will most likely be dominated by inspirals of $\sim10$ \msun\ BHs onto $\sim10^6$\msun\ SMBHs, and could be as high as $\sim 1000$ in 3--5 years within $\sim 3.5$ Gpc.

In this paper, we first in Section~\ref{tf} give an overview of the time-frequency method we proposed and studied previously \citep{wen05a, jon05}.  We then discuss the performance of the method compared to the (semi) coherent search in Section~\ref{perform}. In Section~\ref{info}, we discuss the information about an EMRI event that we can extract using the time-frequency method.  In particular, we introduce a simple decomposition method to decode the directional sensitivity. We also outline the information about the dominant frequency components and their evolution, and the evolution of GW power, that can be derived from the data once a detection is made. We outline possible applications in Section~\ref{discuss}.

\section{A Time-Frequency Method to Detect EMRIs}\label{tf}
In \citep{wen05a, jon05}, we proposed an alternative method for EMRI detection based on the incoherent summation of power in a time-frequency plane.   The t-f power spectrum is produced by dividing the data into segments of equal duration and carrying out a Fast Fourier Transform (FFT) (or alternative spectral decomposition technique) on each.  
In the low-frequency regime, LISA can be regarded as a network of two Michelson interferometers (denoted $I$ and $II$ here), rotated at 45 degrees relative to one another~\citep{curt98}. 
The power spectrum of the detector is defined for each time segment $i$ and frequency bin $k$ as,
\begin{equation}
P(j,k)= 2\frac{|d^{I \, j}_k|^2}{\sigma^2_{Ik}}+2\frac{|d^{II \, j}_k|^2}{\sigma^2_{IIk}},
\label{power}
\end{equation}
where $d^{I,j}_k$ denotes the Fourier amplitude of the $j$-th segment of data from the $I$-th detector, $\sigma^2_{Ik}$ is the expected variance of the noise, $n^I_k$, in the $I$-th detector at frequency bin $k$, assuming the noise is stationary and Gaussian. Our noise model includes the unresolvable background from white dwarf - white dwarf binaries in the usual way \citep{leor04}. In (\ref{power}), the powers of the two data streams have been added directly.  This is optimal based on the maximum likelihood ratio if we have no information about the detector's response and simply assume that the signals included in these two data streams are statistically independent from one other. 

The strategy is then to calculate the power ``density'', $\rho(i,k)$, by computing the average power within a rectangular box centered at each point ($i,k$), 
\begin{equation}
\rho(i,k)= \sum^{n/2}_{a=-n/2} \sum^{l/2}_{b=-l/2} P(i+a,k+b)/m,
\label{rho_tf}
\end{equation}
where $n$, $l$ are the lengths of the box in the time and frequency dimension respectively and $m=n\times l$ is the number of data points contained in the box. To search for a possible signal, we vary the size of the box, and for each choice of $n$ and $l$,  we search for any points at which $\rho(i,k)$ exceeds a threshold determined by a specified false alarm probability set equal for all boxsizes (i.e., we assume an equal chance for any particular clustering of the signal in time and frequency). A detection occurs when one such event happens in any one box size, (see \citep{jon05} for details).  This approach is very similar to the `excess power' technique used in LIGO data analysis \citep{anderson01}, which was designed to search for significant clustering of excess power in the data caused by a source of unknown waveform, but in a given window of time and frequency.  Our method gives a simple estimate of the power density at each point of the time-frequency plane and therefore helps to trace the signal power along the source trajectory.

\section{Performance of the Time-Frequency Method}
\label{perform}

The performance of the time-frequency method can be estimated as follows. When the time-frequency window is ``wrapped tightly around'' most of the signal power, the accumulated signal power increases in proportion to $m$, the size of the window, but the noise power is a $\chi^2$ distribution with $4m$ degree of freedom, the mean and standard deviation of which are $4m$ and $\sqrt{8m}$ respectively. The SNR therefore scales with $\sqrt{m}$ and can be written as
\begin{equation}
SNR_{TF} \sim \frac{\rho^2_m}{\sqrt{8m}}
\end{equation}
where $\rho^2_m$ is the optimal SNR$^2$ if we performed matched filtering on the signal within the time-frequency window in consideration.

The required SNR for a detection via this time-frequency method compared to the fully coherent method can be understood as follows. In the absence of a signal, the output of a matched filter is a Gaussian with zero mean and variance $\rho_m$. The presence of a signal increases the mean to $\rho^2_m$ and therefore the SNR is $\rho_m$. If $m$ is large, the output of the time-frequency search in equation~\ref{rho_tf} in the absence of signal will also be approximately Gaussian, with mean $4$ and variance $8/m$. The presence of a signal enhances the mean by $\rho_m^2/m$. Thus, for a given FAP, the optimal SNR required for detection by the time-frequency search, $\rho_{TF}$, and the optimal SNR required for detection using fully coherent matched filtering, $\rho_c$, are related by
\begin{equation}
\frac{\rho_{TF}}{\rho_c} \sim \frac{(8m)^{1/4}} {{\rho}^{1/2}_c} \ \ \mbox{for a given FAP}.
\end{equation}
The $m$ values for a typical source are around 200--1000.    We note that this is under the approximation of a large window size and assuming we have included roughly all of the signal power. We have also not taken account of the number of trials needed to find the source in either case. In the limit where a signal is uniformly distributed over the whole time-frequency plane, the performance of the time-frequency method is at its worst compared to the coherent method. However, in another limit, not reflected in the preceding equation, for a monochromatic source, the time-frequency method can perform almost as well as the coherent method (provided we somehow know to use a box with $m=1$, and that the coherent method still needs to use the power at the end of the data stream for detection).  In summary, the time-frequency method works better when the signal power is well localized. The actual performance depends on how well we can find a window that encloses most of the signal power to maximize the SNR. 

The detection efficiency of the time-frequency method has been studied in detail in our previous work \cite{jon05},  where only rectangular t-f  windows were used for simplicity.  We used in our simulation a wide range of possible  EMRI signals from the `numerical kludge' approximation \citep{kostas02,gair05,babak06}.  We find that this algorithm is able to detect many different EMRI events out to distances of $1$--$3$ Gpc, depending on the source parameters. In an untargeted search, a typical source can be detected at $2$ Gpc with a detection rate of $60\%
$ at an overall false alarm probability of $1\%$. Lower eccentricity sources, which have less frequency spreading, can be detected as far away as $3$ Gpc with a detection rate of $50\%$ at the same overall FAP. The reach of the search can be extended by increasing the allowed FAP or by using a targeted search. By comparison, the semi-coherent matched filtering algorithm \citep{jon04} can reach $\sim4.5$ Gpc for an overall FAP of $1\%$, but at a presently undetermined detection rate (perhaps $\sim50\%$). Note that the performance of the time-frequency method using a rectangular window is limited by the evolution time scale of the EMRI event.

Broadly speaking, this time frequency search with a rectangular window has better than half the reach of the semi-coherent search \citep{jon04}, but at a tiny fraction of the computational cost. The method can be optimized from using rectangular window by e.g., following the trajectory.   Many image processing methods can be used to improve this shortcoming which we will not discuss in this paper (one possibility is the Hierarchical Algorithm for Clusters and Ridges, which will be described elsewhere~\citep{hacr06}).  Given the simplicity of the technique, we argue that time-frequency  methods could be a valuable first step for detecting the loudest EMRI events in the LISA data stream. 

\section{Extracting Information from the Time-Frequency Spectrogram}\label{info}
There are several issues we must consider when we use the time-frequency method to analyze realistic LISA data. One major challenge is to use the results from the time-frequency method to extract informations about the source which can then be used to reduce the parameter space for a follow up analysis.  The results can help reduce the computational cost for a subsequent coherent or semi-coherent search.

In the following subsections, we will first introduce a simple decomposition method to decode the directional information for EMRIs.  We will then discuss briefly how the outputs can be used to optimize the detection statistic, extract wave information and, to some extent, resolve overlapping sources from distinct directions. We will emphasize EMRI parameter extraction using the t-f method.

\subsection{Directional Information}\label{direction}
\begin{figure}
 \centerline{\includegraphics[height=.3\textheight,width=5in]{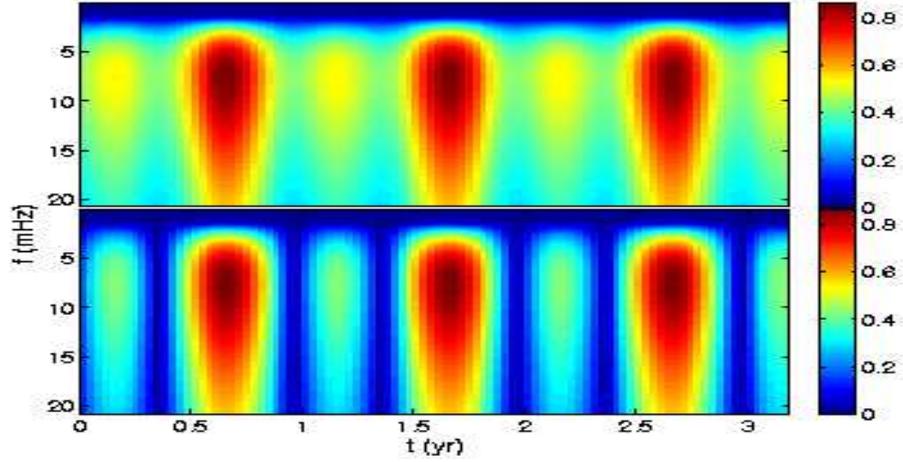}}
  \caption{Singular values $s_1$ (upper panel) and $s_2$ (lower panel) of the response matrix $A$ (Eq.~\ref{svd}) in the time-frequency plane.  The EMRI source is placed at  longitude $\phi =60^o$, and latitude $\theta =57^o$, Ecliptic orbits.  LISA's antenna beam pattern is taken from \cite{curt98}. Higher singular values (redder color) indicate higher directional sensitivities of LISA. }
\label{svd_s}
\end{figure}

LISA can be considered as a network of two detectors in the low-frequency limit and three at high frequencies.  Therefore, we make use of a network analysis carried out for ground-based GW detectors\cite{wen06}. This is summarized as follows. The response of a network of detectors to a source in a given direction is linear. The response in the $k$-th frequency bin and $j$-th segment of time can be written in the frequency domain, after applying appropriate time delays, as 
\begin{equation}
\vec {d}^j_k  = A^j_k \vec{h}^j_k+\vec{n}_k, 
\end{equation}
For the low frequency 2-detector model of LISA at described in \cite{curt98}, we define
\begin{equation}
\vec{d}^j_k=\left ( \begin{array}{c}  d^{j}_{1k}/\sigma_{1k}\\d^{j}_{2k}/\sigma_{2k}\end{array}\right), \,\,    
A^j_k = \left ( \begin{array}{cl} f^{j+}_1/\sigma_{1k} & f^{j\times}_1 /\sigma_{1k}  \\ f^{j+}_2/\sigma_{2k}  & f^{j\times}_2/\sigma_{2k} \end{array}
\right), \,\,    
\vec{h}^j_k =\left ( \begin{array}{c} h^j_{+k}\\ h^j_{\times k}\end{array} \right), \,\,
\vec{n}_k =\left ( \begin{array}{c} n^1_{k}/\sigma_{1k}\\ n^{2}_{ k}/\sigma_{2k}\end{array} \right),
\end{equation}
\label{A_0}
where $f^{j+\times}_i$ are the antenna beam patterns of the $i$-th detector to the $h^j_{+k}$ and $h^j_{\times k}$ polarizations of a GW from a given direction, at time segment $j$. For more than two detectors, the same formalism can be generalized by adding rows in the data and noise vectors and in the response matrix. Using a singular value decomposition, the response matrix $A^j$ can be written as $A^j=usv^T$\cite{NR} where $u$ and $v$ are unitary matrices ($uu^T=I, vv^T=I$),  and $s$ is a diagonal matrix with diagonal values $s_1 \ge s_2$ called singular values.  The signal power from equation~\ref{power} can then be rewritten as a summation of two terms,
\begin{equation}
P(j,k)=|(u^T\vec{d}^{j}_k)_1|^2+|(u^T\vec{d}^{j}_k)_2|^2
\end{equation}
with,
\begin{equation}
 (u^T\vec{d}^{j}_k)_1= s^{j}_1 h'_1+(u^T\vec{n}_k)_1, \ \ (u^T\vec{d}^{j}_k)_2=s^{j}_2h'_2 +(u^T\vec{n}_k)_2, 
\label{svd}
\end{equation}
where $h'_{1,2}$ are the two components of the $\vec{h}'=v^T\vec{h}_k$. We note that each component of $\vec{h}'$ can be measured statistically independently. 
Each term in $P(j,k)$ includes a signal and a noise with the signal proportional to the square of the singular values $s^2_{1,2}$, and a noise term of unity variance.  The singular values $s_1,s_2$ for a given source direction are plotted in Figure~\ref{svd_s} for the LISA configuration described in \cite{curt98}. It is clear that for most of LISA's orbit (regions of red-color), the two directional sensitivities are comparable and therefore the power summation in equation (\ref{power}) is optimal.  However, for a significant fraction of time (region of blue color), one sensitivity can be much less than the other and in some parts of the time-frequency plane, the sensitivity can be nearly zero.

The significance of this is that the summation method in equation~\ref{power} can be further optimized by summing up only terms with large singular values to optimize the SNRs. In other words, we use the data from one or both of the time-frequency planes shown in Figure~\ref{svd_s} only when they have good sensitivity to the sky direction under consideration. It is clear that the improvement in detection efficiency for this simple 2-detector approximation is limited. However, the situation will be different for configurations accounting for the rotation of the LISA constellation (which encodes more directional information), and at higher frequencies when LISA becomes effectively a 3-detector network. For a 3-detector network, at least one null-stream (a term with zero singular values) that is a linear combination of the three data streams can be constructed.  The detection statistics should improve more significantly in this case. The null-stream can also be used to localize the source and as a consistency check to run in parallel with the detection method \cite{wen06}.

A direction specific search is required for such a decomposition to work.  Based on Figure~\ref{svd_s} and previous studies on the lower limit of LISA's angular resolution \cite{curt98} and work by Pai et al.\cite{pai06}, the angular resolution of LISA will be quite rough for a time-frequency method. A search grid to cover all sky directions can be as coarse as tens of degrees and the search over multiple directions is therefore not too computationally intensive. 

\subsection{Constraints on Parameter Space}\label{param}
Three types of information can be obtained from the time-frequency method once a detection is made --- (1) $f_n(t)$,  the time evolution of the dominant frequency components in the time-frequency plane, where $n$ indicates the number of harmonic this frequency is of its fundamental one.  An example of this can be seen in the bright ``typical'' case described in our previous work \cite{wen05a, jon05}. (2) $\dot{f}_n(t)$, the time derivatives of the dominant frequencies.  This can be calculated,  e.g., by finite differencing. (3) $<|h(t,f_n)|^2>$, the signal power in the dominant frequencies,  where the brackets $<>$ indicate the average over the time-frequency windows that yield SNRs above the threshold and along the trajectory.  The signal power $|h(t,f_n)|$ can be extracted by inversion of equation (\ref{svd}). 
The measurement uncertainty can be calculated based on the singular values.  Small singular values should be discarded using the standard treatment for an ill-conditioned matrix~\cite{NR}.  
The continuous signal power, $<|h(t,f_n)|^2>$,  can then be estimated from its discrete representation.

A full analytical solution to the dynamical evolution of these three quantities is not yet available for EMRIs. We will therefore make use of the post-Newtonian(PN) treatment described in \cite{leor04} to illustrate how  the time-frequency method might be used to put constraints on the system parameters.  First we must understand the dependence of the dominant frequency components, $f_n(t)$, on the system parameters. For a circular orbit around a non-spinning black hole, the dominant GW power output is at twice the orbital frequency. However, when the central black hole has spin and the orbit is eccentric, the GW output is significantly colored by precession effects. The majority of the GW energy is emitted close to periapse, when the object is on a nearly circular ``whirl'' orbit~\cite{GK2002,steve06}. The dominant GW emission will therefore be at the effective ``whirl'' frequency and its harmonics. We are currently developing expressions for the dependence of the whirl frequency on system parameters. In the PN formalism~\cite{leor04}, we can write
\begin{equation}
f_n(t)=n\nu(t)+\dot {\gamma} (t)/\pi = F_1(\nu(t), e(t), \mu, M, S, \lambda),
\label{fc}
\label{eqn1}
\end{equation}
where $\nu(t)$ is the radial orbital frequency at time $t$, and $\dot {\gamma} (t)$ is the rate of pericenter precession (given by equation (29) of \cite{leor04}).  The value of $n$ that determines the dominant frequency component (i.e., the ``whirl'' frequency) depends non-trivially on the eccentricity and spin. The dominant frequencies, $f_n(t)$, are therefore functions of $\nu(t)$, eccentricity $e(t)$, reduced mass $\mu$, total mass $M$, magnitude of the SMBH spin $S$, and $\lambda$, the inclination of the orbit with respect to the spin direction of the SMBH (assumed fixed in \cite{leor04}).

The time derivatives, $\dot{f}_n (t)$, can also be related to those of the system parameters
\begin{eqnarray}
\dot{f}_n (t) = F_2(\nu(t), e(t), \mu, M, S, \lambda), 
\label{eqn2}
\end{eqnarray}
where,
\begin{equation}
F_2 = \dot{\nu} \frac{\partial F_1}{\partial \nu} + \dot {e}\frac{\partial F_1}{\partial e}+\dot{\lambda} \frac{\partial F_1}{\partial \lambda}.
\end{equation}
The time derivatives $\dot{\nu}$, $\dot{\lambda}$ and $\dot{e}$ are results of gravitational radiation reaction.  PN expressions for them are given in equations (28)-(30) of \cite{leor04} where $\lambda$ is assumed to be fixed.

\begin{figure}
 \centerline{\includegraphics[height=.3\textheight,width=5in]{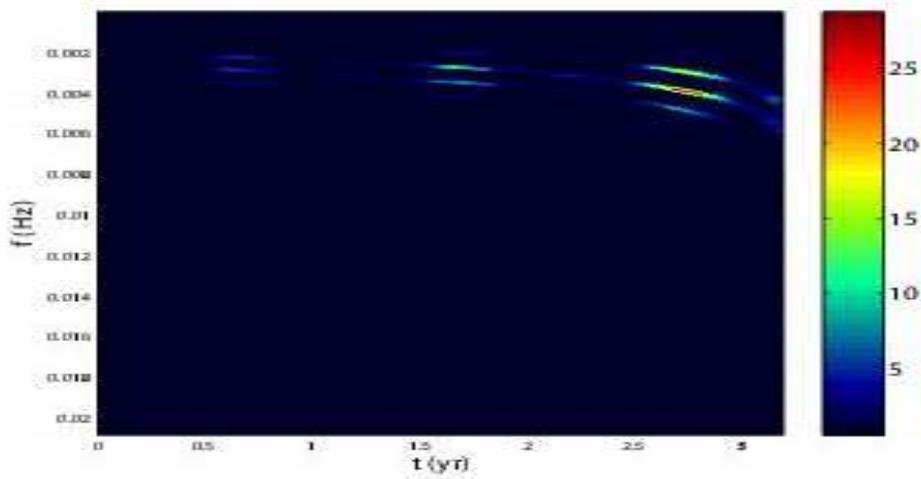}}
\caption{The t-f power density for a typical EMRI source at $d=0.5$ Gpc (see text). Harmonic structures of signal frequencies and yearly amplitude modulation caused by LISA's directional sensitivity are evident (cf Fig.~\ref{svd_s}). Optimistically, we expect $< \sim 3$ such events in three years. }
\label{fig_tf}
\end{figure}

\begin{figure}
 \centerline{\includegraphics[height=.3\textheight,width=5in]{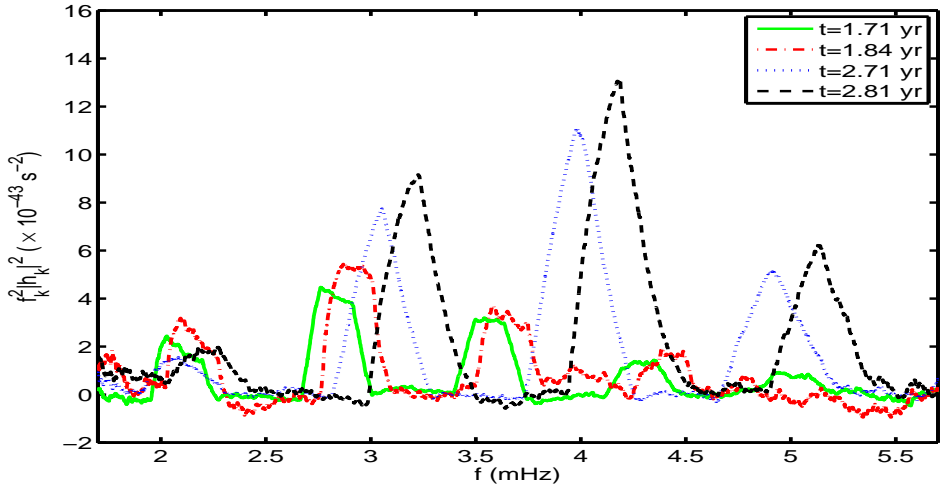}}
\caption{The normalized energy power $f^2 |h_f|^2$ vs frequency at four different times for the same EMRI source of Fig.~\ref{fig_tf} at $d=0.5$ Gpc. The 3rd --6th harmonics of the radial frequency can be identified at $t=1.71$ yr (solid curve) and at $t=1.84$ yr (dashdot line).  The four peaks at later times of $t=2.71$ yr (dotted line) and of $t=2.81$ yr (dashed line) are very close to the 2--5th harmonics of the radial frequency.  }
\label{fig_harm}
\end{figure}

A relation between a GW  frequency, its derivative and the signal power $|h(t,f_n)|^2$ at that frequency is
\begin{equation}
<|h(t,f_n)|^2> =\frac{G}{\pi^2 c^2 D^2}\dot{E}_n/(\dot{f}_nf^2_n) =F_3 (\nu(t), e(t), \mu, M, S, \lambda, D).
\label{eqn3}
\end{equation}
where $D$ is the distance to the source, $F_3=G/(\pi^2c^2D^2)\dot {E}_n/(F_1F_2)$. Under the quadrupole approximation, the energy power radiated at the $n$-th harmonics can be written as~\cite{peters} 
\begin{eqnarray}
\dot {E}_n=\frac{32}{5}\frac{G^{7/3}}{c^5}\mu^2M^{4/3} (2\pi\nu (t))^{10/3}g(n,e(t)).
\label{eqn4}
\end{eqnarray}
As an illustration, we  show in Fig.~\ref{fig_tf} the time-frequency power density for a ``typical'' EMRI source of $10+10^6$ \msun\  at $d=0.5$ Gpc (same as Fig.~1, left panel of \cite{wen05a}).  Trajectories of frequency evolution vs time and harmonic structures are apparent.  The weighted average of the quantity $f^2_k |h_k|^2$ for the same source are then calculated for each point on the t-f plane using Eq.~\ref{svd}, assuming that we have obtained directional information.   In Fig.~\ref{fig_harm}, we plot $f^2_k |h_k|^2$ vs frequencies at four different times.  The signal power distributions among frequency harmonics and their evolution are clearly observable.  Data points of signal frequencies, their derivatives ($f_n$, $\dot f_n$)  and the fundamental frequency $\nu$,  can be easily measured from the figure. Once the  harmonic structures are identified, eccentricities can be estimated from the relative power in different harmonics (from function g(n,e) in Eq.~\ref{eqn4}) and $\mu M^{2/3}/D$ can be estimated. A quantitative study for this approach is under way.

In summary equations~(\ref{eqn1}), (\ref{eqn2}) and (\ref{eqn3}) provide an illustration of how constraints on the parameter space and information on $e(t), \nu(t)$ etc. can be derived from a t-f map. If we assume that we have detected $N_h$ harmonics and that along each trajectory we have $N_t$ independent data points above the detection threshold, then in principle $3N_tN_h$ data points can be used to constrain the $2N+5$ unknowns $\nu(t), e(t), \mu, M, S, \lambda$, and $D$ If the SNR is high enough, these parameters can then be calculated using, e.g., the least squares method. 


\section{Discussion}\label{discuss}
We have reviewed a time-frequency method that can be used to detect bright EMRI sources. We have discussed the performance of this method --- previous studies have shown that this method can be effective in detection of EMRIs at distances of up to 1--3 Gpc depending on the system parameters. We have proposed a recipe based on singular value decomposition to decode the directional information about EMRI sources. We have presented a method to constrain the parameter space using the PN evolution equations given in \cite{leor04}. The parameter space can be constrained using equations (\ref{eqn1}), (\ref{eqn2}), and (\ref{eqn3}). For a detection of $N_h$ harmonics with $N_t$ data points on each trajectory (assuming all trajectories are generated by the same EMRI), there are in principle $3N_hN_t$ data points that can be used to constrain the $2N_t+5$ unknowns $e(t), \nu(t), \mu, M, S, D$ and $\lambda$ for the PN expressions in \cite{leor04}. Therefore, in this simplified model, it is possible, in principle, that the total number of observed data points from the time-frequency method can be larger than the number of unknowns and the equations can be solved.  A quantitative study is underway and results will be presented in a follow-up paper.  

Another unsolved issue is the problem of confusion caused by the WD-WD binaries in the LISA data. Three different types of information obtained from the time-frequency method can be used to distinguish them from EMRIs. One is the frequency binsize which gives the maximum SNR, the other is the shape of trajectories and the third is the directional information.  The maximum SNR of a signal can be obtained only in the correct direction and with box sizes smaller than the signal spread. WD-WD binary signals are expected to be single-bin (since the Doppler shift is negligible for our frequency bin sizes) and the trajectories should exhibit very little time evolution. 
 
At the end, the time-frequency method provides directly measurements of EMRI in dominant frequencies and their time derivatives.  These information can be used to map out the space-time geometry of the SMBHs\cite{ryan97,hughes06}. This will be the subject of different studies (e.g., \cite{chao06}).  
 
\begin{theacknowledgments}
This work was supported in part by the Alexander von Humboldt Foundation's Sofja Kovalevskaja Programme (funded by the German Federal Ministry of Education and Research) and by St.Catharine's College, Cambridge.

\end{theacknowledgments}



\bibliographystyle{aipproc}   




\end{document}